\documentclass[sigconf]{acmart}
\renewcommand\footnotetextcopyrightpermission[1]{} % removes footnote with conference information in first column
\usepackage[utf8]{inputenc}
\usepackage{natbib}
\usepackage{graphicx}
\usepackage[caption=false]{subfig}
\usepackage{multirow}
\usepackage{multicol}
\usepackage{array}
\usepackage{xcolor}
\usepackage{float}
\usepackage[normalem]{ulem}
\usepackage{listings}
\usepackage{makecell}
\usepackage{hyperref}
\usepackage{xspace}

\newcommand{\dcdb}{DCDB\xspace}
\newcommand{\daf}{Wintermute\xspace}
\newcommand{\cmuc}{CooLMUC-3\xspace}
\newcommand{\lrz}{LRZ\xspace}

\title[Operational Data Analytics with \dcdb \daf]{\dcdb \daf: Enabling Online and Holistic Operational Data Analytics on HPC Systems}
\author{Alessio Netti}
\email{alessio.netti@lrz.de}
\affiliation{Leibniz Supercomputing Centre}
\affiliation{Technical University of Munich}
\author{Micha M\"uller}
\email{micha.mueller@lrz.de}
\affiliation{Leibniz Supercomputing Centre}
\affiliation{Technical University of Munich}
\author{Carla Guillen}
\email{carla.guillen@lrz.de}
\affiliation{Leibniz Supercomputing Centre}
\author{Michael Ott}
\email{michael.ott@lrz.de}
\affiliation{Leibniz Supercomputing Centre}
\author{Daniele Tafani}
\email{daniele.tafani@lrz.de}
\affiliation{Leibniz Supercomputing Centre}
\author{Gence Ozer}
\email{gence.ozer@tum.de}
\affiliation{Technical University of Munich}
\author{Martin Schulz}
\email{schulzm@in.tum.de}
\affiliation{Technical University of Munich}
\affiliation{Leibniz Supercomputing Centre}

\acmConference[HPDC '20]{The International ACM Symposium on High-Performance Parallel and Distributed Computing}{June 23-26, 2020}{Stockholm, Sweden}

\copyrightyear{2020}
\acmYear{2020}
\setcopyright{acmcopyright}\acmConference[HPDC '20]{Proceedings of the 29th International Symposium on High-Performance Parallel and Distributed Computing}{June 23--26, 2020}{Stockholm, Sweden}
\acmBooktitle{Proceedings of the 29th International Symposium on High-Performance Parallel and Distributed Computing (HPDC '20), June 23--26, 2020, Stockholm, Sweden}
\acmPrice{15.00}
\acmDOI{10.1145/3369583.3392674}
\acmISBN{978-1-4503-7052-3/20/06}

\begin{document}

\begin{abstract}
As we approach the exascale era, the size and complexity of HPC systems continues to increase, raising concerns about their manageability and sustainability. For this reason, more and more HPC centers are experimenting with fine-grained monitoring coupled with Operational Data Analytics (ODA) to optimize efficiency and effectiveness of system operations. However, while monitoring is a common reality in HPC, there is no well-stated and comprehensive list of requirements, nor matching frameworks, to support holistic and online ODA. This leads to insular ad-hoc solutions, each addressing only specific aspects of the problem.

In this paper we propose \daf, a novel generic framework to enable online ODA on large-scale HPC installations. Its design is based on the results of a literature survey of common operational requirements. We implement \daf on top of the holistic \dcdb monitoring system, offering a large variety of configuration options to accommodate the varying requirements of ODA applications. Moreover, \daf is based on a set of logical abstractions to ease the configuration of models at a large scale and maximize code re-use. We highlight \daf's flexibility through a series of practical case studies, each targeting a different aspect of the management of HPC systems, and then demonstrate the small resource footprint of our implementation.
\end{abstract}


\maketitle
\renewcommand{\shortauthors}{Netti et al.}

\section{Introduction}
\label{section:introduction}

The computational requirements of modern scientific research grow steadily, and \emph{High-Performance Computing} (HPC) systems are designed accordingly with ever-increasing scale and parallelism, taking us to the brink of the \emph{exascale} era, with systems capable of computing a billion billion operations per second. However, the enormous and intricate complexity of these systems resulting from the adoption of heterogeneous architectures, novel cooling systems, as well as complex management software to support modern applications and workflows, makes both their deployment and production use a challenge. Further, concerns about excessive power consumption~\cite{villa2014scaling} and high failure rates~\cite{cappello2014toward} question their feasibility. To counter these challenges, it will be more important than ever before to treat HPC machines as dynamic, complex systems themselves whose efficiency and effectiveness must be proactively and continuously monitored, analyzed and improved~\cite{babaoglu2018cognified}: in other words, the operation of \emph{all} available resources must be orchestrated \emph{at all times} in a systematic, holistic and automated manner. This does not only include compute resources, but also memory, network, I/O and infrastructure resources among others.

Analyzing the operation of an HPC system to gain insight into its behavior is the purpose of \emph{Operational Data Analytics} (ODA), as defined by Bourassa et al.~\cite{Bourassa:2019:ODA:3339186.3339210}. ODA is driven by the large amounts of data produced by monitoring frameworks, which capture and store data at fine granularity from a large number of sensors in hardware and software components. These sensors are located across the entire HPC facility, from the infrastructure down to the compute node level, and can be used to infer knowledge about system behavior, enabling system \emph{control} via the implementation of a proactive control loop. If implemented in an online fashion, such a control loop allows to automate the tuning of system knobs that otherwise would be fixed or manually set, allowing for significant improvements in terms of energy efficiency, system reliability and total cost of ownership, among other aspects~\cite{bautista2019collecting}.

Monitoring and ODA are therefore two key aspects in the design of future HPC systems. However, while monitoring is an established reality in most supercomputing centers~\cite{ahlgren2018large}, ODA is still far from it: many experimental solutions address individual issues ranging from node resiliency to infrastructure management and energy efficiency, but they are insular and rarely adopted in production. As shown in a survey conducted by the Energy Efficient HPC Working Group, in fact, most HPC sites use ODA in a visualization context, relying on general-purpose frameworks and without automating system control~\cite{eehpc2019}. The main reason for this lies in the absence of comprehensive frameworks founded on well-defined requirements, which could then enable the wider adoption of ODA approaches on HPC systems and facilities. Due to the wide variety of needs inherent to specific ODA techniques, a framework of this kind must be designed to cope with the extreme volumes of data associated with monitoring a wide and diverse set of sensors, as well as the tight latency and overhead constraints of real-time system control.

\paragraph{Related Work}
\label{section:relatedwork}
 
The problem of enabling ODA on HPC systems in a generic, holistic way is still an open research question. The \emph{Lightweight Distributed Metric Service} (LDMS)~\cite{agelastos2014lightweight} has been recently enhanced to support ODA features on top of standard HPC monitoring~\cite{izadpanah2018integrating}. However, due to its pull-based architecture, it is not suitable for in-band, fine-grained ODA applications that require live data with minimal overhead and latency. Moreover, LDMS currently lacks global configuration abstractions to simplify the instantiation of models on large-scale HPC systems.

The \emph{Examon}~\cite{beneventi2017continuous} framework is shown to be suitable for ODA applications, being based on the MQTT protocol~\cite{locke2010mq} and thus compatible with out-of-band tools such as \emph{Apache Spark}. However, this reliance on the use of external tools to process data results in a complex software stack that needs to be tuned \emph{ad-hoc} for each specific use case, as well as in non-optimal data retrieval performance. The \emph{OMNI}~\cite{bautista2019collecting} framework has a similar architecture, but is more oriented towards visualization of data and misses the abstractions necessary for control. The \emph{GUIDE}~\cite{vazhkudai2017guide} framework combines monitoring and ODA features, but it is mostly log-oriented and the semantics of its data analytics features are not clear. \emph{Elastic Stack}\footnote{\url{https://www.elastic.co/products/}} supports the post-processing of data ingested from external sources, thus enabling data analytics for monitoring frameworks such as \emph{Ganglia}~\cite{massie2004ganglia}. The analysis, however, is centralized at the server level, which limits scalability for large HPC installations.

Many other tools propose basic applications of ODA specifically tailored for HPC and implement simple feedback loops between the monitoring component and the resource manager (e.g., the \emph{Energy-Aware Runtime} (EAR)~\cite{corbalan2019ear} or \emph{IBM LoadLeveler}~\cite{auweter2014case}). Similarly, tools like \emph{SPar}~\cite{griebler2018service} provide user-friendly interfaces for runtime tuning. These efforts, however, tackle specific issues of resource management in HPC systems, and customization for other purposes is not trivial. Further, due to the lack of coordination mechanisms, using multiple systems of this kind concurrently to address multiple resources can even be counterproductive. The \emph{Global Extensible Open Power Manager} (GEOPM)~\cite{eastep2017global} provides a plugin-oriented and extensible interface for resource and power management in HPC systems, but its monitoring capabilities are limited. Alongside the open-source solutions discussed above, there are also many commercial and closed-source products, such as \emph{Zenoss}\footnote{\url{https://www.zenoss.com/}} or \emph{Splunk}\footnote{\url{https://www.splunk.com/}}, offering extensive data analytics capabilities. However, these products are often designed solely for loosely-coupled data center environments and are not suitable for use in HPC centers.

In summary, to the best of our knowledge, there is no generic and comprehensive solution addressing the problem of online ODA on HPC systems, and hence we need a novel approach to tackle ODA on next-generation supercomputers. 

\paragraph{Contributions}
In this paper we tackle this research gap in the ODA field and present \emph{\daf}, a novel framework to enable online and holistic operational data analytics on HPC systems, capable of processing data and taking decisions at any level of the system. We designed \daf following an extensive literature review and requirements analysis, as well as based on previous experiences in single-point ODA solutions at our supercomputing center, which allowed us to identify the main functional and operational requirements for a generalized ODA framework. \daf's workflow accommodates most real-world ODA applications, while at the same time its small resource footprint renders it suitable for applications in which overhead and latency are critical. We implement \daf within the \emph{Data Center Data Base} (\dcdb) monitoring system~\cite{netti2019dcdb}. Our contributions are the following:

\begin{itemize}
    \item We propose a taxonomy of ODA techniques for HPC systems based on a literature survey, classify them according to their modes of operation, and extract common requirements.
    \item We introduce the \daf framework, which enables the analysis of data and control at all levels in the hierarchy of an HPC system, and implement it within \dcdb.
    \item We introduce an approach, called the \emph{block system}, to aid in the instantiation of ODA models on large-scale HPC systems using a tree representation of the sensor space.
    \item We demonstrate the applicability and scalability of \daf through a series of case studies carried out on an HPC system at the Leibniz Supercomputing Center (LRZ).
\end{itemize}

\paragraph{Organization}
The paper is organized as follows. In Section~\ref{section:requirements} we outline the design requirements for our framework. In Section~\ref{section:architecture} we describe the architecture of \daf, alongside its integration in \dcdb in Section~\ref{section:deployment}. In Section~\ref{section:units} we discuss the block system, while in Section~\ref{section:usecases} we present a series of case studies we implemented. In Section~\ref{section:conclusions} we conclude the paper.
\section{Analysis of Requirements}
\label{section:requirements}

First we present the use case analysis for the design of the \daf framework, following a literature survey and extracting common functional and operational requirements.

\subsection{Uses of Operational Data Analytics}
\label{subsection:uses}

Even though ODA techniques are emerging for managing many aspects of HPC systems, they have not been systematically classified and typical functional requirements are still not clear, to the best of our knowledge. This, however, is a fundamental prerequisite for the design of a generic framework: for this reason we propose a non-exhaustive \emph{taxonomy}, depicted in Figure~\ref{requirements:taxonomy}, identifying the most common use cases associated with ODA on HPC systems. This list is based on recent and relevant works, and reflects the trends in ODA at most HPC sites, including the experiences at \lrz. In particular, we identify the following main usage scenarios:

\begin{itemize}
    \item \textbf{Infrastructure Management}: optimizing the operation of infrastructure and facility-wide systems (e.g., cooling or power distribution), as well as adapting to environmental changes~\cite{conficoni2015energy, jiang2019fine, grant2015overtime, jha2018characterizing}.
    \item \textbf{Scheduling and Allocation}: improving the placement of user jobs on an HPC system by supplying additional information (e.g., system energy budgets, thermal limits or I/O features) to the scheduler~\cite{bash2007cool,imes2018energy,sirbu2016power, verma2008power}.
    \item \textbf{Prediction of Job Features}: using heuristic or learning techniques to predict the duration of user jobs and their submission patterns, improving the effectiveness of scheduling policies and reducing queuing  times~\cite{galleguillos2017data,naghshnejad2018adaptive,matsunaga2010use,emeras2015evalix,ejarque2010semantic}.
    \item \textbf{Application Fingerprinting}: optimizing management decisions by predicting the behavior of user jobs and correlating this to historical data to characterize features such as power consumption and network usage~\cite{ates2018taxonomist,gallo2015analysis,zhang2012hpc,wyatt2018prionn,mckenna2016machine}.
    \item \textbf{Fault Detection}: detecting and predicting anomalous states in hardware and software components to improve the resiliency of HPC systems, preventing in turn unmasked failures and other catastrophic events~\cite{guan2013adaptive,tuncer2018online,sirbu2016towards,shaykhislamov2018approach}.
    \item \textbf{Runtime Tuning}: predicting the behavior of applications and components in compute nodes for dynamic tuning using system knobs (e.g., CPU frequency)~\cite{eastep2017global,corbalan2019ear,wang2017modular,Lin2016}.
\end{itemize}

\subsection{Taxonomy of Operational Data Analytics}

The list of use cases above demonstrates that ODA is needed at all levels of an HPC system, as well as at different time scales; all techniques, however, rely directly on monitoring data and some applications, such as those associated with job analysis, may further require additional data (e.g., job id or wall time). Based on our observations, we derive four classes of ODA techniques, according to the type of data they use and their mode of operation. On one hand, we identify two types of data sources:

\begin{itemize}
    \item \textbf{In-band}: data sampled and consumed within a specific component in an HPC system, usually a compute node. Techniques using such data sources often operate at a fine temporal scale (i.e., greater than 1Hz) and require low analysis overhead and latency in gathering data.
    \item \textbf{Out-of-band}: data potentially coming from any available source in the system, including historical or asynchronous facility data. In a few cases, job-related data may be used as well. For techniques using this type of data, operation often has to be at coarse scale (e.g., in the order of minutes) and must be explicitly synchronized (e.g., through time-stamps), but latency and overhead are less of a concern.
\end{itemize}

On the other hand, we group ODA techniques according to the two following modes of operation:

\begin{itemize}
    \item \textbf{Online}: continuous operation, producing an output resembling a time series, which can then be re-used to drive management decisions and thus produce a feedback loop.
    \item \textbf{On-demand}: operation triggered at specific times (e.g., job submission) to steer management decisions that require certain information about the system's status.
\end{itemize}

\begin{figure}[t]
 \centering
 \includegraphics[width=0.43\textwidth,trim={0 0 0 0}, clip=true]{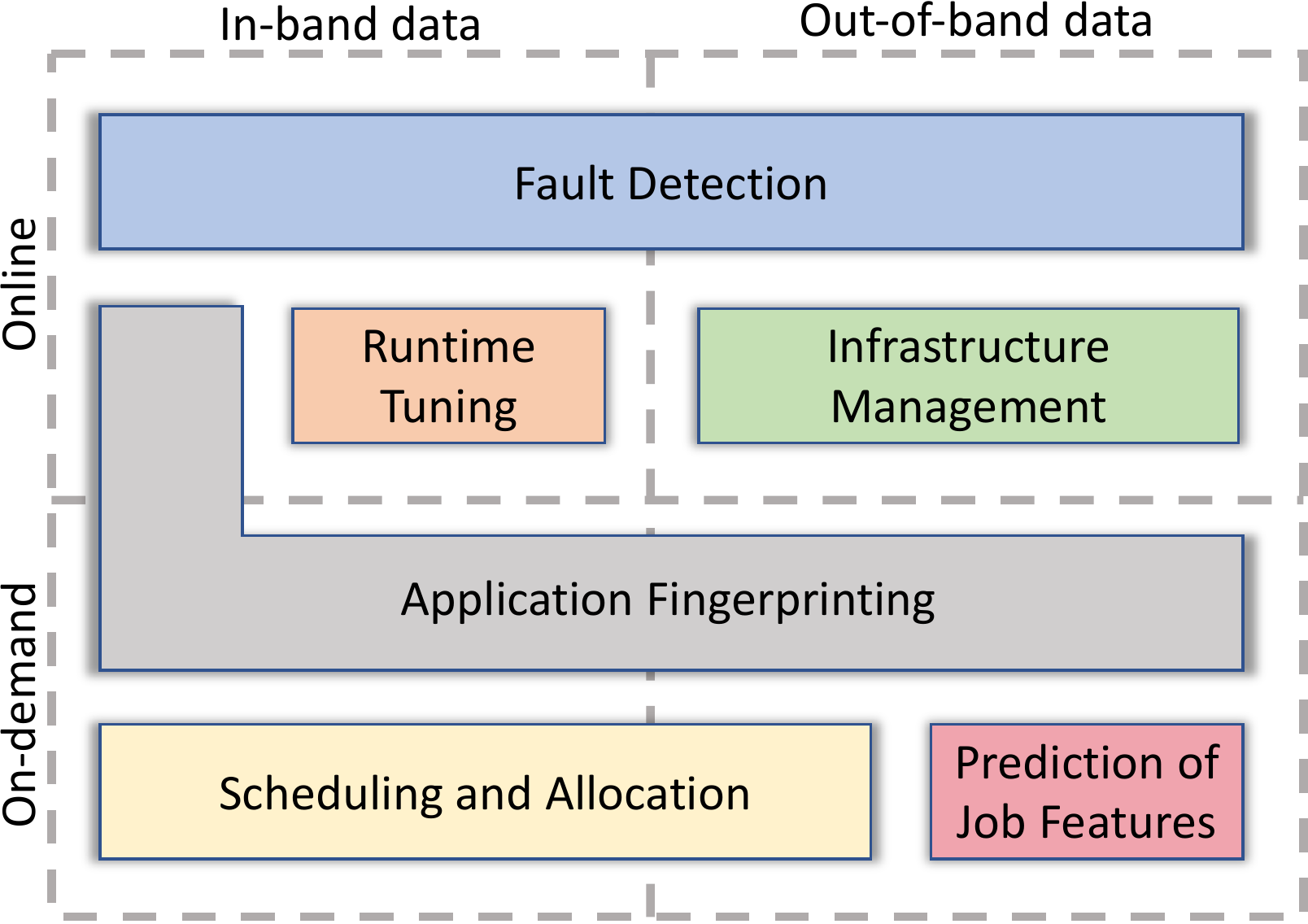}
 \caption{A taxonomy of common ODA applications in HPC systems. Use cases spanning multiple classes can be employed in either mode depending on their specific scenarios.}
 \label{requirements:taxonomy}
 \end{figure}

Using these characteristics we can classify the use cases presented in Section~\ref{subsection:uses} as shown in Figure~\ref{requirements:taxonomy}, leading us to a taxonomy, which we use in the following to guide our design of \daf.

\subsection{Functional Requirements}

In light of our taxonomy of ODA techniques, we extract a series of functional and operational requirements that must be taken into account when designing any generic online ODA framework for HPC systems, including our \daf framework:

\begin{itemize}
    \item \textbf{Holism}: an ODA framework must provide a holistic view of an HPC system's sensor space, exposing available data in the way that is most fitting to the current scenario. An \emph{in-band} ODA model will benefit from in-memory processing of local sensor data, for optimal latency and overhead. Conversely, an \emph{out-of-band} model performing coarse-grained analysis may require large amounts of data (e.g., historical) that cannot be maintained within local memory and thus must be fetched from remote storage.
    \item \textbf{Flexibility}: both \emph{online} and \emph{on-demand} operations must be supported to address the necessities of different techniques driven by the various components of an HPC system.
    \item \textbf{Scalability}: ODA models must be able to scale up to thousands of inputs and very fine time scales. At the same time, an ODA framework must exhibit a light resource footprint, to not interfere with HPC applications when used in-band.
    \item \textbf{Abstraction}: manual configuration of ODA is prohibitive when a large amount of independent models (e.g., one per CPU core of an HPC system) must be deployed together, both in-band and out-of-band. For this reason, abstraction constructs are necessary to simplify and automate the configuration of ODA models at scale.
    \item \textbf{Modularity}: as knobs and sensors in HPC systems are often controlled via a set of common and pre-defined protocols (e.g., IPMI or SNMP), an ODA framework must be modular and able to integrate a wide range of external interfaces. Further, as ODA techniques often rely on similar processing steps (e.g., regression), it must allow the pipelining of several analysis and control stages so as to maximize code re-use. 
\end{itemize}

Careful system design is needed to address the requirements above. Most recent efforts in ODA for HPC systems (e.g., Examon~\cite{beneventi2017continuous}) rely on the use of tools such as \emph{Apache Spark} or \emph{Elastic Stack}: these allow for scalability and holism, but not the needed flexibility. Further, the support of in-band models is severely limited, addressing only part of the problem. To overcome these limitations and to address all requirements, we require a tight integration of ODA with the components of an holistic monitoring system specifically targeting HPC: monitoring interacts with most HPC components addressed by ODA naturally, providing efficient and convenient interfaces for control as well as streaming data access.

\begin{figure}[t]
 \centering
 \includegraphics[width=0.47\textwidth,trim={0 0 0 0}, clip=true]{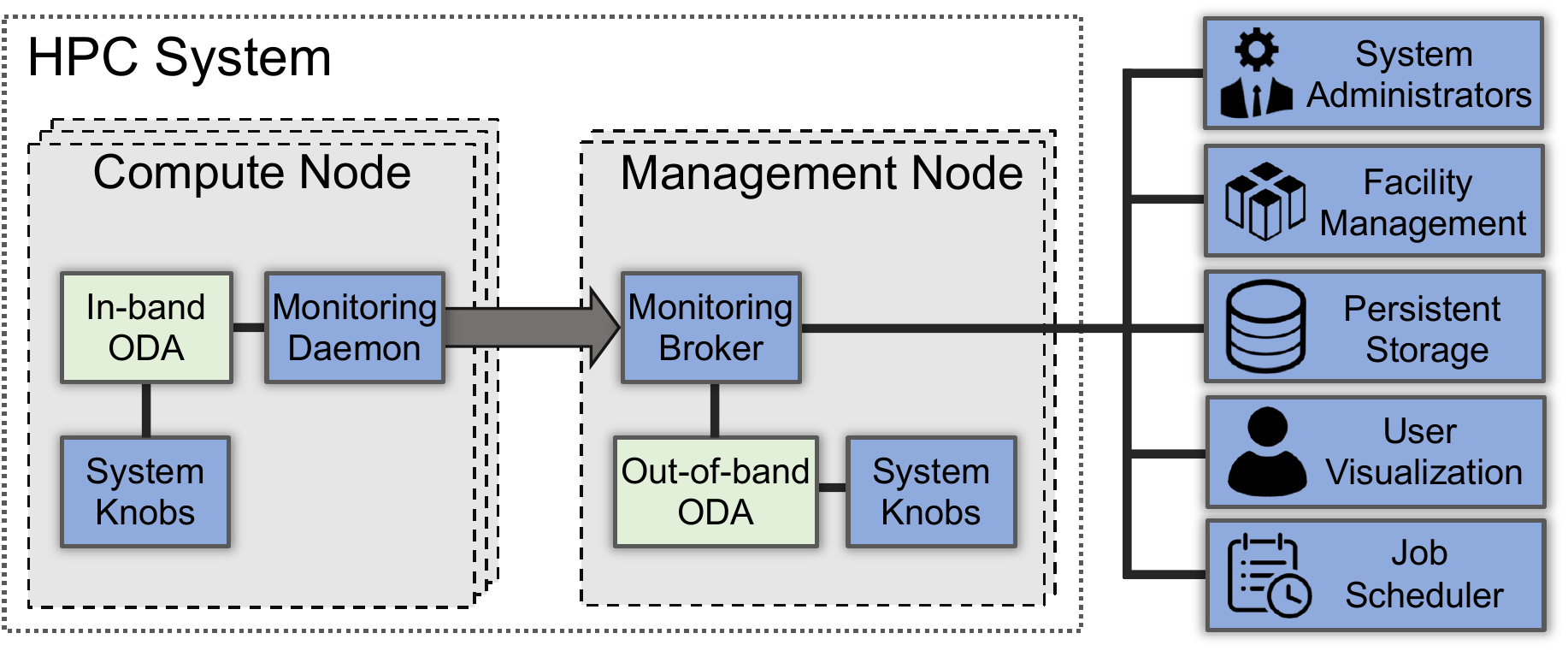}
 \caption{A high-level overview of the suggested architecture for an online ODA framework integrated in a monitoring system, showing the main components and actors involved.}
 \label{requirements:archsketch}
 \end{figure}

Figure~\ref{requirements:archsketch} shows how ODA can be implemented within an existing monitoring system: integration with monitoring daemons in compute nodes enables in-band operation, close to where data is sampled, whereas management nodes are used for out-of-band operation. In the latter case, ODA can interact with a monitoring data broker, gaining access to streamed cluster-wide data, as well as remote persistent storage. This approach covers all use cases and requirements laid out in this section.
\section{Architecture of \daf}
\label{section:architecture}

Following the design guidelines laid out in Section~\ref{section:requirements} we introduce \daf, a novel ODA framework driven by our requirements analysis. We first provide an overview of \daf's architecture, and then describe in detail the components comprising it.

\subsection{Architecture Overview}

\begin{figure}[t]
\centering
\includegraphics[width=0.42\textwidth,trim={0 0 0 0}, clip=true]{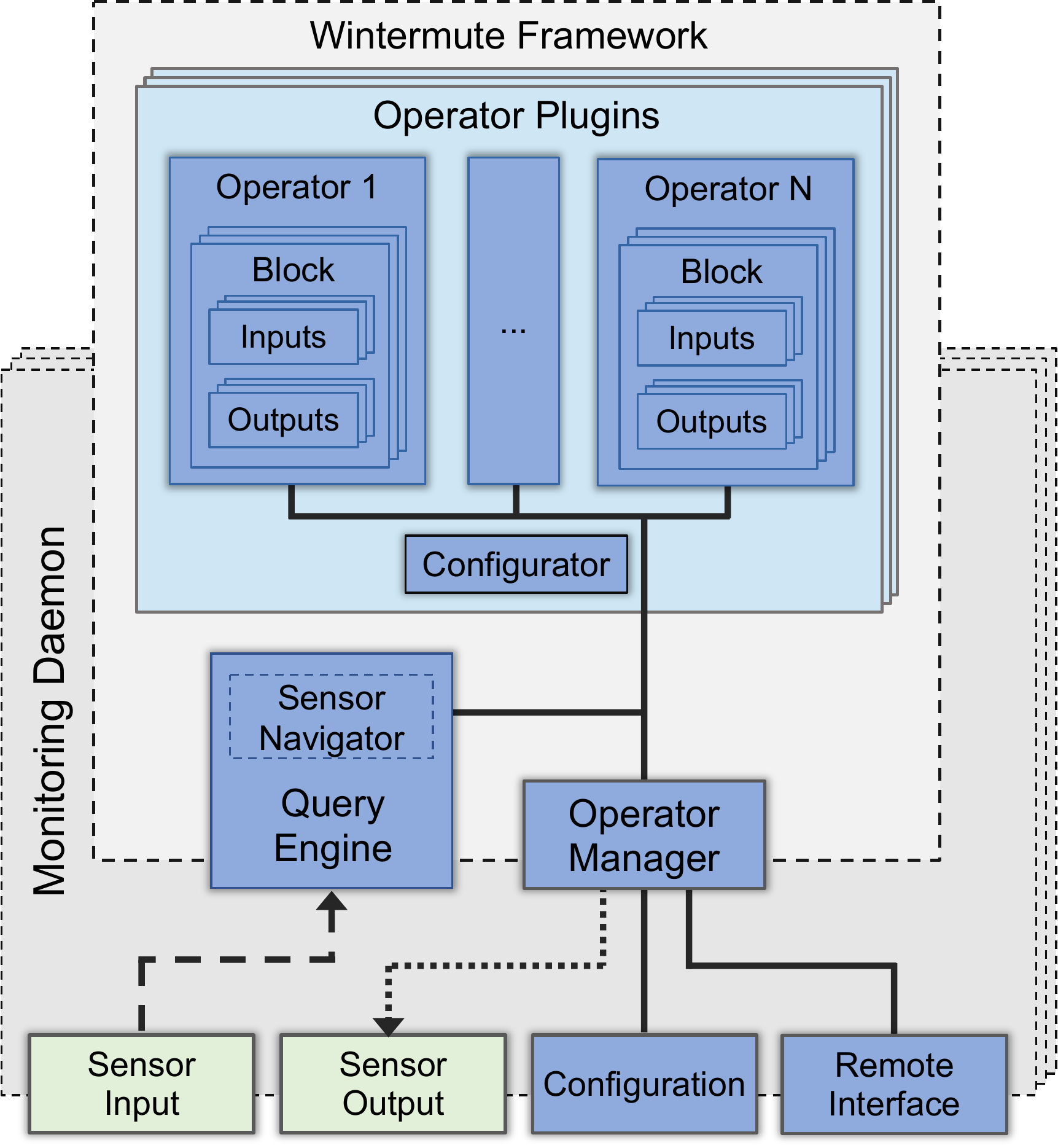}
\caption{Architecture of \daf. We abstract from its integration in a monitoring system, and only show the external components with which it interacts. Components linked by dashed or dotted lines are integration-dependent.}
\label{architecture:arch}
\end{figure}

\daf provides an ODA framework with generic interfaces and is designed in a way that it can be integrated into any HPC monitoring system as an additional software component. In Figure~\ref{architecture:arch} we show its modular architecture: it is based on \emph{operator plugins} supplying analysis capabilities, which follow an agnostic code interface and are used to instantiate \emph{operators}. Operators represent the actual computational entities performing all ODA tasks asynchronously, by relying on a flexible local thread pool. Each operator works on a set of \emph{blocks}, which are container data structures representing physical components (e.g., compute nodes or racks) or logical entities (e.g., user jobs) in an HPC system: a block has a set of sensors that are used as inputs for the analysis (\emph{input} sensors), as well as a set of outputs, which store the results of the ODA operation and are, again in the form of sensors, to be consumed by the monitoring system or by other operators (\emph{output} sensors). In our terminology, a sensor defines an atomic monitoring entity (e.g., power, temperature, CPU counter or ODA output) that captures system information. Each sensor reading is identified by a numerical value (in a unit given by the sensor's definition) and a time-stamp. We will return to the concept of blocks with greater detail in Section~\ref{section:units}.

Operator plugins are supported by two central components, the \emph{query engine} and the \emph{operator manager}, which provide input data to operators and expose their output respectively. These are designed to isolate the plugins from the location in which they are instantiated, meaning that a plugin can be deployed to the different locations of a monitoring system (or different monitoring systems altogether) without alterations. The last set of components, depicted at the bottom of Figure~\ref{architecture:arch}, belongs to the monitoring system in which \daf is integrated: the \emph{sensor input} and \emph{sensor output} components describe the interfaces through which \daf obtains \emph{sensor} data and exposes analysis results respectively. The \emph{configuration} component is responsible for initialization and will grant \daf access to its designated configuration files, which are indicated in the global monitoring system's configuration. The \emph{remote interface} component, finally, represents the interface exposed by the monitoring daemon, through which \daf can in turn expose its remote control and data retrieval features. While a RESTful API is our preferred interface type, \daf is not dependent on this choice and can easily be adapted to work with other interface types (e.g., remote shells).

\subsection{Components of the Architecture}

In the following we describe the core components that compose the \daf architecture, as well as their interactions with the surrounding monitoring system.

\paragraph{Operator Manager}
The operator manager is the central entity responsible for loading requested \daf operator plugins, exposing the associated configuration files to them and managing their life cycle. As such, it is the main interface between \daf and the monitoring system and allows users to specify which sensors to read. Additionally, the operator manager acts as a front-end for all remote interface requests (e.g., via a RESTful API), exposing available actions implemented within the framework. For example, these requests can be used to start, stop or load plugins dynamically, as well as trigger specific actions on a per-plugin basis (e.g., training a machine learning model) and retrieve recent sensor data.

\paragraph{Query Engine}
The query engine is a \emph{singleton} component that exposes the space of available sensors to operator plugins. In particular, it gives access to a \emph{sensor navigator} object, which maintains a tree-like representation of the current sensor space using the block system described in Section~\ref{section:units}, allowing \daf plugins to discover which sensors are available and where in the hierarchical structure they stand. The query engine's uniform interface enables queries based on sensor names and time-stamp ranges. Access to low-level sensor data structures is achieved by means of a callback function, which is set at startup by the monitoring entity in which the \daf framework is running. Access to job data and other metadata can be enabled by setting similar callback functions.

\paragraph{Operator Plugins}
\label{subsection:operatorplugins}
Operator plugins implement the specific logic to perform analysis processes of a certain kind, complying to the \daf plugin interface. \emph{Operator plugins} perform analysis by taking as input sensor data alone. \emph{Job operator plugins} are an extension of normal operator plugins which can also use job-related data (e.g., user id or node list), producing output associated to a specific job. Plugins consist of the following main internal components:
 
\begin{itemize}
    \item \textbf{Operator}: operators are objects performing the required analysis tasks. Each operator has assigned a set of blocks, each referencing a set of input and output sensor objects. Whenever computation is invoked for an operator, it will iterate through its blocks and perform an analysis for each of them, querying the respective input sensors through the query engine, processing the obtained readings, and storing the result in the output sensors. When performing analysis for a certain block, access to the operator's other blocks is allowed for correlation purposes.
    \item \textbf{Configurator}: a configurator is responsible for reading a plugin's own configuration file, exposed by the operator manager, and instantiating operators together with their blocks: the process to generate the latter is controlled by a series of template-based constructs that allow users to easily instantiate a large number of blocks (e.g., one per CPU core in a large-scale HPC cluster), each with their own unique sets of input and output sensors. This mechanism is discussed in detail in Section~\ref{subsection:unitconfiguration}, and more resources about the configuration can be found in \dcdb \daf's GitLab.
\end{itemize}

\section{\daf Integration into \dcdb}
\label{section:deployment}

As discussed in Section~\ref{section:architecture}, \daf must be tightly coupled with a corresponding monitoring framework tasked with providing sensor data and allowing transport and storage of results. In our concrete implementation, we integrate \daf with the \dcdb monitoring framework~\cite{netti2019dcdb}, which is used at \lrz: we first briefly describe \dcdb, followed by the integration of \daf into it, and finally present the resulting workflow and associated options. Like \dcdb, \daf is implemented in C++11, and all source code is freely available under the GNU GPL license via its GitLab repository\footnote{\label{note1}https://dcdb.it}. It also includes a series of end-to-end examples, demonstrating the simplicity of \daf's configuration process. Due to the abstract and generic nature of \daf's architecture, our C++11 implementation can be re-used and integrated into any other monitoring system with little effort.

\subsection{Architecture of \dcdb}
\dcdb is a holistic solution for continuous monitoring in HPC systems~\cite{netti2019dcdb}. It comprises several components in order to achieve a distributed and scalable architecture, which is summarized in Figure~\ref{architecture:flow}: \emph{Pushers} perform the sampling of sensors on monitored components, using a plugin-based architecture that allows to easily add new data sources. All collected data is sent via the MQTT protocol~\cite{locke2010mq} to \emph{Collect Agents}, which act as data brokers and forward the data to a \emph{Storage Backend}, currently implemented using \emph{Apache Cassandra}. Alongside a series of interfaces for visualizing and retrieving data from Storage Backends, \dcdb also exposes a RESTful API for control in every component, as well as sensor caches for fast in-memory access to recent readings.

\subsection{Workflow of \daf}
\begin{figure}[t]
\centering
\includegraphics[width=0.45\textwidth,trim={0 0 0 0}, clip=true]{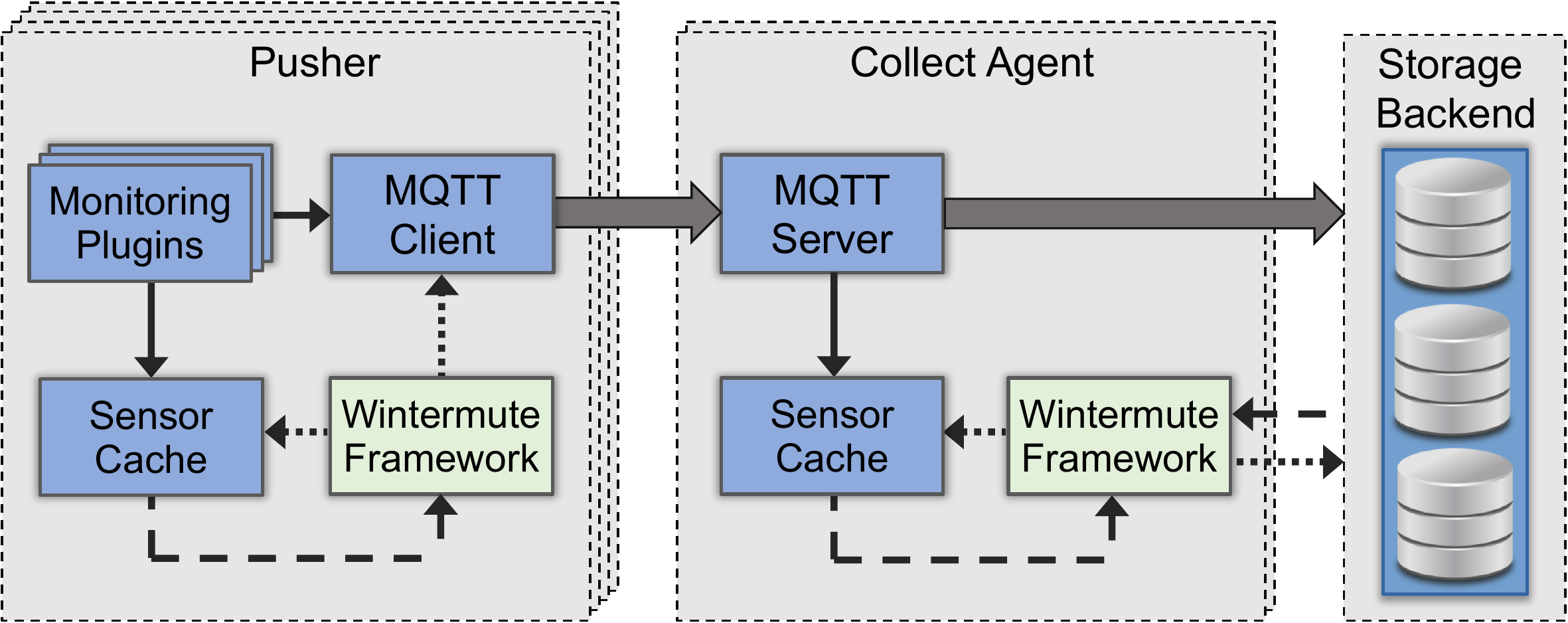}
\caption{High-level overview of the architecture of \dcdb, highlighting the \daf framework's integration in components and the data flow.}
\label{architecture:flow}
\end{figure}

\daf is included in Pushers and Collect Agents as an additional plugin-based software component that enhances them by supplying ODA capabilities, as described in Section~\ref{section:architecture}. Figure~\ref{architecture:flow} shows the integration of \daf in the existing \dcdb architecture: it has access to all resources in a Pusher or Collect Agent, including sensor caches, RESTful APIs and data output methods (i.e., MQTT or Storage Backend). The arrows directed in and out of the \daf components define the inputs and outputs for sensor data in each location. In the following we discuss the resulting available options that allow the configuration of \daf's workflow to accommodate the use cases laid out in Section~\ref{section:requirements}.

\paragraph{Operator Location}
As \daf is included in Pushers and Collect Agents, operators can be instantiated in both locations by loading the appropriate plugins. In a Collect Agent, access to the entire system's sensor space is available. If possible, data is retrieved from the local sensor cache or otherwise queried from the Storage Backend, to which the outputs of operators are also written. This location is optimal for system or infrastructure-level analysis and feedback loops. In a Pusher, on the other hand, operators have only access to locally sampled sensors and their sensor cache data. This location is optimal for runtime models requiring data liveness, low latency and horizontal scalability. For example, a regression operator used to predict power consumption for CPU frequency tuning~\cite{ozer2019towards} can be deployed in a Pusher so as to leverage in-memory processing for minimal latency.

In both scenarios, the query engine gives higher priority to data in the local sensor caches, which is faster to retrieve compared to querying the Storage Backend. Moreover, queries can be performed in two modes, affecting how the caches are accessed: in the first, \emph{relative} time-stamps are supplied as an offset against the most recent reading, and the cache view to be returned can be computed in $O(1)$ time. In the second, \emph{absolute} time-stamps are used, resulting in a binary search with $O(log(N))$ time complexity.

\paragraph{Operational Modes}
In the native \daf implementation, operators can be configured to work in two different ways depending on their requirements. In \emph{online} mode, an operator is invoked at regular time intervals, resulting in continuous analysis and thus producing time series-like sensor data as its output. This is ideal for applications such as fault detection or runtime tuning. In \emph{on-demand} mode, on the other hand, an operator's capabilities must be explicitly invoked via the RESTful API, by querying a specific block. Output data is propagated only as a response to the RESTful request. This mode is ideal for scheduling applications, which can be triggered at arbitrary times. For example, a resource manager could contact an on-demand fault detection operator at scheduling time~\cite{tuncer2018online} to determine the current status of each idle compute node and thus optimize allocation decisions.

\paragraph{Block Management}
When using the online mode, the blocks of a single operator can be arranged with respect to the underlying model: as \emph{sequential}, all blocks share the same operator, and are processed sequentially at each computation interval to avoid race conditions; as \emph{parallel}, one distinct operator is created for each block, allowing us to parallelize computation and improve scalability. For example, an application fingerprinting operator~\cite{ates2018taxonomist} deployed in a Collect Agent with a large number of blocks, one per compute node in an HPC cluster, could make use of the \emph{parallel} option for optimal ODA performance.

\paragraph{Analysis Pipelines}
As the output data produced by online operators shares the same format and is identical to all other sensor data in \dcdb, operators can use the output of other operators as input. This, in turn, allows us to create \emph{pipelines}, in which the multiple stages of a complex analysis are divided among several operators. This can be used to split computational load between multiple locations (e.g., Pusher and Collect Agent) or to achieve complex analyses with few, general-purpose plugins. Furthermore, this method allows us to implement feedback loops in an HPC system, via \emph{control} operators at the end of a pipeline that use processed data to tune knobs. For example, an operator which estimates the optimal inlet cooling water temperature for an HPC cluster based on sensor data~\cite{conficoni2015energy} could feed its predictions to a second operator, devoted to issuing SNMP requests to tune the cooling system.

These options provide effective ways to implement a lightweight and reliable data analytics infrastructure: for example, on-demand operation can be used to minimize overhead and data volume; careful planning of operator pipelines, instead, can reduce redundancy in processing of sensor data, by allowing multiple operators to consume the data produced by a single one.

\section{The Block System}
\label{section:units}

With more and more data sources to tap into, navigating the space of available sensors in a monitored HPC system and configuring ODA models at scale becomes difficult and error-prone~\cite{gimenez2017scrubjay}. This calls for a structured sensor specification system with the following requirements: a) it must simplify the navigation of large sensor spaces, with millions of entries; b) it must allow to derive the hierarchical relationships existing between sensors; c) it must provide template-like constructs to simplify sensor specifications for ODA models. Here, we introduce the set of abstractions we implemented in \daf to address these issues.

\subsection{The Sensor Tree}
In \daf we model the sensor space as a hierarchical \emph{sensor tree}, an example of which is depicted in Figure~\ref{architecture:sensortree}. We assume that the keys (or \emph{topics}, similarly to the MQTT standard) used to identify sensors are forward slash-separated strings similar to file system paths, expressing their physical or logical placement in an HPC system. The following is an example of a sensor topic:
\begin{lstlisting}
/rack4/chassis2/server3/power
\end{lstlisting}

The last segment of a topic is the name of the sensor itself, and the preceding path elements express its placement in the system. This representation can be exploited to construct a tree, in which each internal node is a system component (e.g., a compute node or a rack) and each leaf is a sensor. The constructed tree then supplies a comprehensive view of the monitored system's structure, as well as a natural way to correlate hierarchically-related sensors (e.g., the sensors of a compute node and those of the rack it belongs to). 

The structure of the sensor tree is analogous to a file system: components of the HPC system represented by internal tree nodes can be seen as \emph{directories} whereas the sensors themselves corresponding to leaves are akin to \emph{files}. This approach has been already employed in efforts such as \emph{Perftrack}~\cite{knapp2007perftrack}, proving its effectiveness. Here, we extend it for the purpose of ODA model configuration.

The effectiveness of this representation depends on the level of detail expressed by the hierarchy of topics, and the responsibility for devising such a hierarchy lies on system administrators and designers. Some HPC centers might also employ monitoring systems with naming conventions different from the file system-like one we discuss here. This is taken into account in our implementation, which is not dependent on a specific naming scheme, but also supports the definition of arbitrary hierarchy schemes supplied as lists of regular expressions each identifying a separate tree level.

\subsection{Blocks and Block Templates}
\begin{figure}[t]
 \centering
 \includegraphics[width=0.43\textwidth,trim={0 0 0 0}, clip=true]{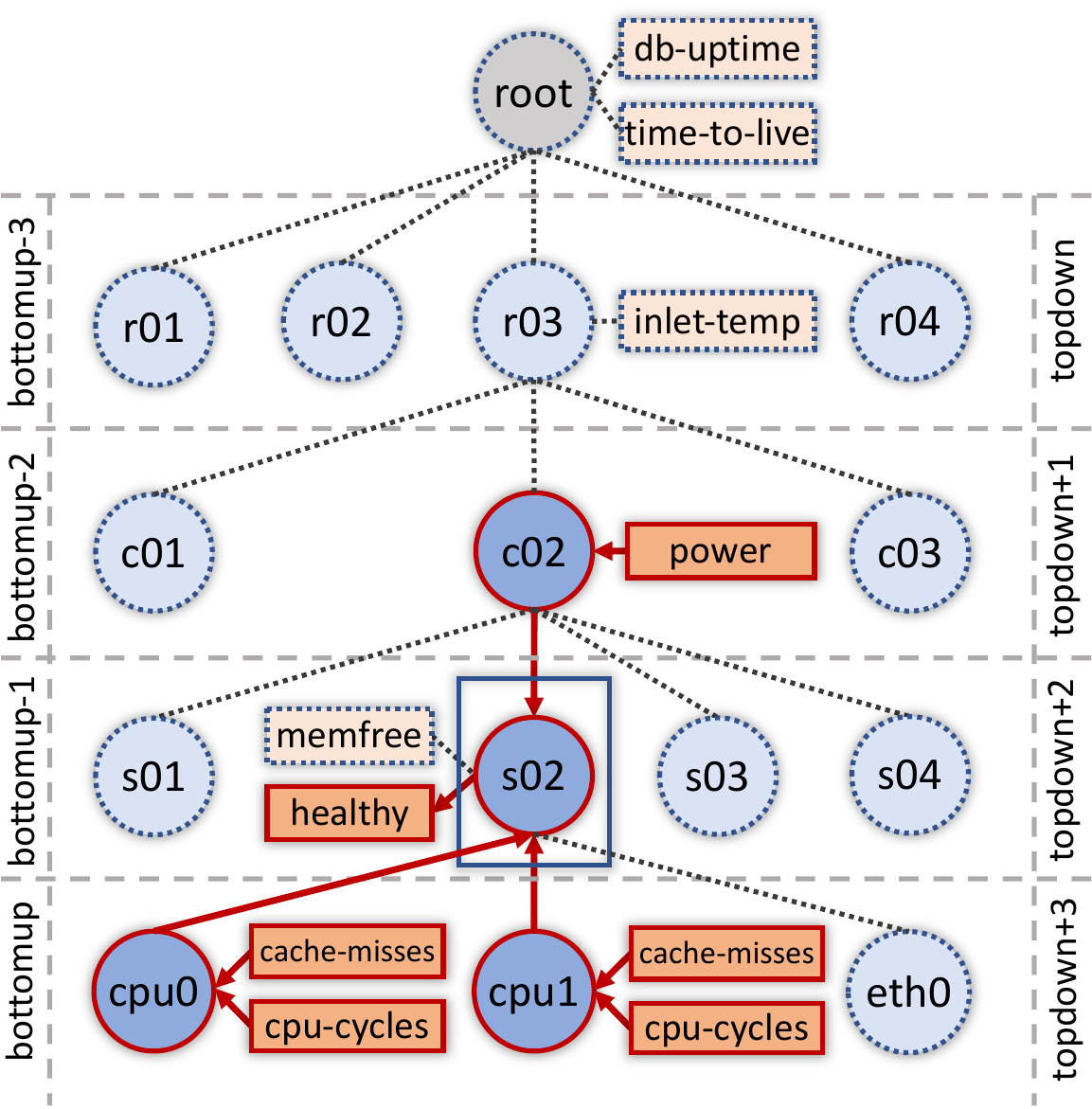}
 \caption{The sensor tree of an HPC system, and a \daf block. Circles represent internal tree nodes, while sensors are represented by rectangles. Red, non-dashed lines highlight the nodes and sensors belonging to the block, while other nodes are collapsed for convenience.}
 \label{architecture:sensortree}
 \end{figure}

In \daf, \emph{blocks} are data structures that act as atomic containers on which analysis computations are performed. A block represents directly a node in the sensor tree, from which it takes its \emph{name}. Then, a block references a set of \emph{input} and \emph{output} sensor topics: the output sensors are used to deliver analysis results, and are leaves of the node the block represents. Input sensors, which provide the data for the analysis, can either be leaves of that same node, or belong to any other node in the sensor tree connected by an ascending or descending path to it. Figure~\ref{architecture:sensortree} shows a generic example for a block, named \emph{s02}, a compute node in an HPC system. In this example, the block has the output sensor \emph{healthy}, and a series of input sensors: the \emph{cycles} and \emph{cache misses} counters of the CPUs in the compute node, plus the \emph{power} sensor of the chassis it belongs to. Combined with its input and output sensors, a block corresponds to a \emph{sub-tree} in the sensor tree. 

While blocks can be defined by specifying actual sensor topics as inputs and outputs, they may also be defined in a generic way via \emph{templates}: here, sensors are referenced via their position in the sensor tree, expressing only their last topic segment and omitting the components to which they belong (preceding topic path), which are replaced by a tree level (\emph{vertical} navigation) and a filter (\emph{horizontal} navigation). The set of sensor topics matched by a sensor expression is its \emph{domain} in the tree. Further, a particular block can be instantiated from a template by specifying a node in the sensor tree (i.e., its name), thus creating a binding: each sensor expression is then replaced with a sensor topic from its domain that is hierarchically-related to the block's node. Since multiple topics may satisfy this, one expression can produce multiple actual topics. Conversely, if no topic satisfies it, the block cannot be built. 

Recalling the similarity between the sensor tree and a file system, describing sensors through sensor expressions can be interpreted as defining files using \emph{relative paths}: these paths can match multiple points in the file system, and they are fully resolved in function of the \emph{current working directory}, whose analogous in this case is the name of the block. The main difference between the two is that the tree level of sensors in sensor expressions is defined with an absolute level, whereas for relative file system paths it is defined as a relative offset with respect to the current working directory.

\subsection{Template Instantiation}
\label{subsection:patterninstantiation}

The example block shown in Figure~\ref{architecture:sensortree} can be built from a generic template using the following sensor expressions:
\begin{lstlisting}
input:
    <topdown+1>power
    <bottomup, filter cpu>cpu-cycles
    <bottomup, filter cpu>cache-misses
output:
    <bottomup-1>healthy
\end{lstlisting}

In sensor expressions, the \emph{topdown} and \emph{bottomup} keywords drive the \emph{vertical} navigation and indicate the highest and lowest level in the tree, respectively; the root node of the sensor tree is excluded from this representation, and other levels can be reached through relative offsets. The \emph{filter} keyword defines the \emph{horizontal} navigation and is used to filter the set of topics that the expression matches, within its tree level. In this example, the block's name is set to \emph{/r03/c02/s02/}, which identifies an HPC compute node. Once this is set, the rest of the block is resolved: the \emph{power} expression is resolved as \emph{/r03/c02/power}, since it specifies that the sensor should be one level below the highest tree level, at \emph{c02}. Conversely, the \emph{cpu-cycles} and \emph{cache-misses} expressions are on the lowest level, with two nodes (\emph{cpu0} and \emph{cpu1}) belonging to their domains. As such, sensors from both of them are added to the block. As the \emph{healthy} output sensor expression lies at the same level as \emph{s02}, it is simply resolved as \emph{/r03/c02/s02/healthy}.

\subsection{Configuring Blocks in \daf}
\label{subsection:unitconfiguration}

Templates are used in \daf's plugin configurators to instantiate the blocks operators work on, as explained in Section~\ref{subsection:operatorplugins}. In detail, the block generation process works in the following steps, starting from a block template defined in a configuration file: 

\begin{enumerate}
    \item based on the current sensor tree, the set of topics matching the output sensors' expressions (their domain) is computed;
    \item one block is created for each retrieved node in the domain;
    \item for each block, its set of input and output sensors is resolved according to the domains of the respective expressions.
\end{enumerate}

On top of block-level outputs, users may also define a set of operator-level outputs that can, for example, store the average error of a model applied to a set of blocks. Recalling the example of template in Section~\ref{subsection:patterninstantiation}, applying the configuration algorithm described above will result in as many blocks as compute nodes in the HPC system (e.g., \emph{/r03/c02/s[01-04]/}). This demonstrates how the block system enables instantiation of thousands of independent ODA models in a large-scale HPC system, each with its own set of sensors, by using only a small configuration file. Moreover, configurations are independent from the location to which a model is deployed (e.g., Pusher or Collect Agent), as the blocks are resolved automatically from the available sensor tree. It should be noted, however, that a block template is not guaranteed to be portable across HPC systems with different sensor hierarchies. For example, if we wanted to port the template in Section~\ref{subsection:patterninstantiation} to a system whose hierarchy has an additional level close to the root of the tree, with block names such as \emph{/i01/r03/c02/s02/}, the expression \emph{<topdown+1>power} must become \emph{<topdown+2>power}.

\section{Case Studies}
\label{section:usecases}

In this section we present several case studies showing the capabilities of \daf, alongside an analysis of the required resources and overheads. These case studies were not chosen for their novelty and represent (on purpose) typical ODA techniques in the literature, so as to show \daf's flexibility and suitability for large-scale HPC installations: without a framework such as \daf, carrying out these case studies would be difficult and would require the development of a substantial amount of dedicated, non-reusable code to retrieve sensor data, place and control the analysis and to expose its output. All experiments described in this section were carried out on the \cmuc system at \lrz\footnote{\url{https://doku.lrz.de/display/PUBLIC/CoolMUC-3}}. This cluster is composed of 148 compute nodes, each equipped with a 64-cores Intel Xeon Phi 7210-F \emph{Knights Landing} CPU, 96GB of RAM and an Intel \emph{Omni Path Architecture} (OPA) interconnect. \dcdb runs continuously on this system in production, with Pushers in compute nodes sampling data from the \emph{Perfevent}, \emph{SysFS}, \emph{ProcFS} and \emph{OPA} plugins and with a single Collect Agent forwarding the data to a dedicated Storage Backend.

\subsection{Power Consumption Prediction}
\label{subsection:powerusecase}

\begin{figure}[t]
\centering
\captionsetup[subfigure]{}
\subfloat[Time series of the real and predicted power values.]{
\includegraphics[width=0.40\textwidth,trim={35 0 0 0}, clip=false]{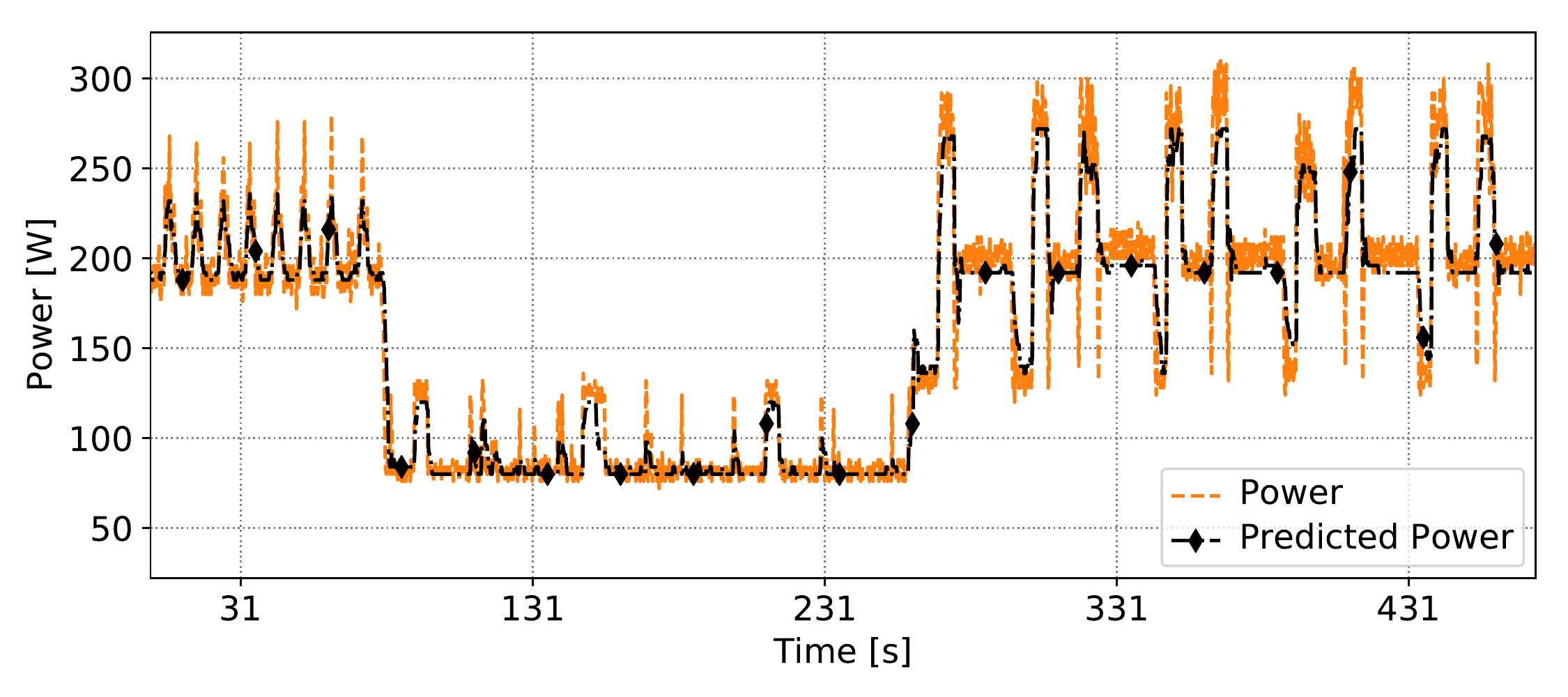}
  } \\
\subfloat[Relative error of the predicted power values.]{
\includegraphics[width=0.47\textwidth,trim={0 0 0 0}, clip=true]{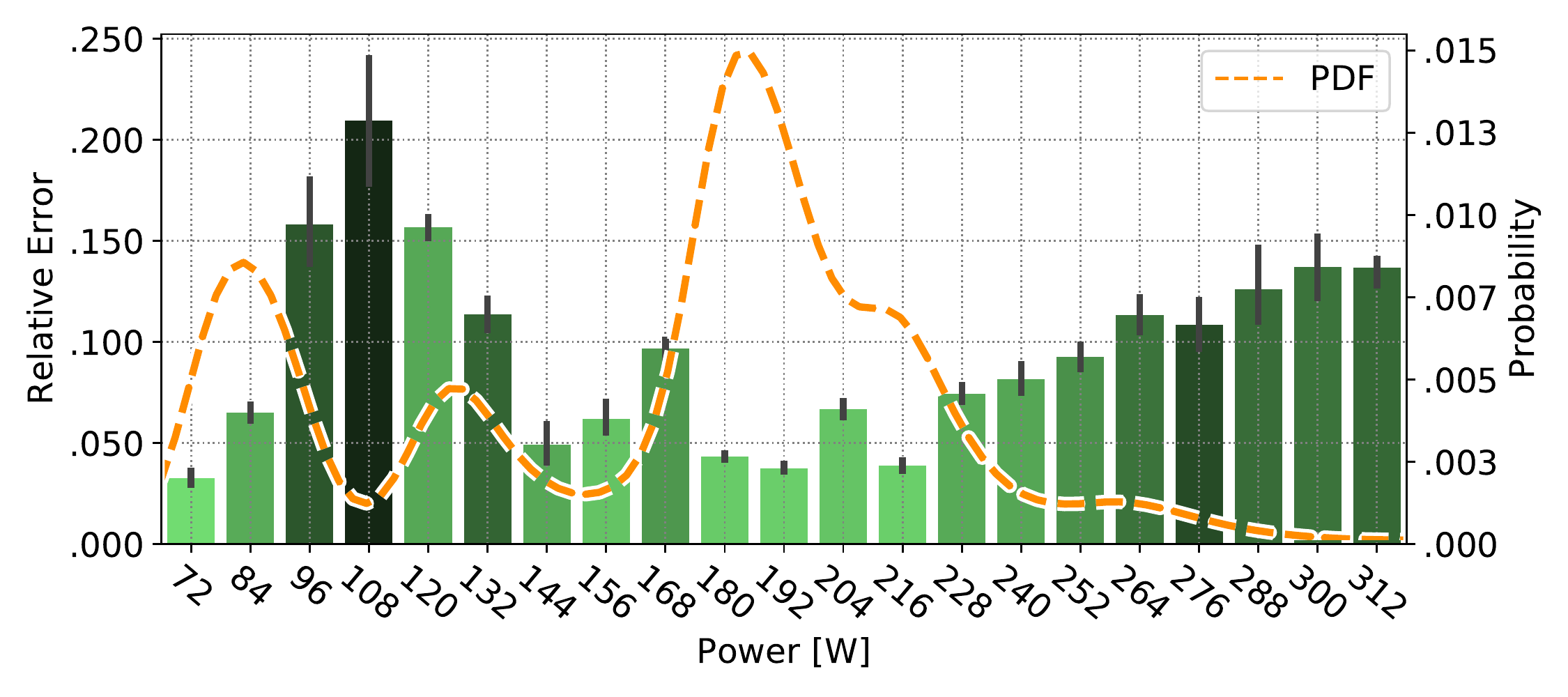}
}
\caption{Performance of our power consumption prediction model in terms of time series behavior and relative error. Average relative error is 6.2\%.}
\label{results:regression}
\end{figure}

The first study shows the use of \daf for predicting the power consumption of a compute node (precisely, overall node power measured at the power supply) in \cmuc, which can be used to steer online control decisions in the power and runtime systems. An example of this is DVFS CPU frequency tuning, which can be exploited in an automated way to save significant amounts of energy without sacrificing application performance. In this scenario data is collected in-band, at a fine time scale, and is immediately re-used for control purposes. The model represents an online implementation of the one proposed by Ozer et al.~\cite{ozer2019towards}. 

\subsubsection{Configuration} In a Pusher, we instantiate a single operator from a plugin, called \emph{Regressor}, which implements a generic random forest-based online regression model. Its input data consists of a set of performance metrics and sensors, and both sampling and regression operate at a 250ms interval. The plugin, which is based on the OpenCV library\footnote{\url{https://opencv.org}}, works in the following way: at each computation interval, for each input sensor of a certain block, a series of statistical features (e.g., mean or standard deviation) are computed from its recent readings. These features are then combined to form a feature vector, which is fed into the random forest model to perform regression and output a sensor prediction for the next 250ms. 
The training of the model, which is shared by all blocks of an operator, is performed automatically: feature vectors are accumulated in memory until a certain training set size is reached, alongside the responses from the sensor to be predicted. In this case, the responses come from the power sensor, with the model set to predict its value in the next 250ms. With the Pusher running, we execute the \emph{Kripke}, \emph{AMG}, \emph{Nekbone} and \emph{LAMMPS} proxy HPC applications from the CORAL-2\footnote{\url{https://asc.llnl.gov/coral-2-benchmarks}} suite, with as many threads as physical cores, while the regression operator builds its training set. Here the operator has only one block, corresponding to the compute node, and the training set size is set to 30,000. Once training is complete, we evaluate the regression with new \dcdb data.

\subsubsection{Results} Figure~\ref{results:regression} summarizes the results of the model. It shows a small excerpt from the time series of the real and predicted power sensors: we see that the time series of the predicted power consumption follows the measured time series closely, capturing status changes and periodic behaviors before they occur. However, the predicted time series fails to capture some short spikes or oscillations in power consumption, and presents itself like a \emph{smoothed} version of the measured one. These events are difficult to predict, as they are usually related to the CPU's power management policy, which may exhibit short-term spikes for throughput improvement (e.g., \emph{Turbo} mode on Intel CPUs) or may be related to electrical or sensor noise. 
The phenomena described above are even more apparent in Figure~\ref{results:regression}b, which shows the average relative prediction error for each real power band, together with the fitted probability density function (PDF) of the latter. It can be seen that prediction is worse for higher power values; as it can be observed from the PDF, these values represent a minimal portion of the distribution, and have negligible impact on the overall error. Moreover, this imbalance in the distribution translates directly to an imbalance in the training set of the model, which does not have enough data to capture this type of behavior. Similarly, some low-power states that are relatively rare are not predicted well by the model. However, in the regions of the distribution where most of the data concentrates, error is always close to 5\%, proving the model's effectiveness. 

We obtained comparable average relative error values when sampling and predicting power consumption at a time interval of 125ms (10.4\%) and 500ms (6.7\%). In the work by Ozer et al.~\cite{ozer2019towards}, the offline validation of this approach shows comparable results to the ones presented here, proving its generality. While specialized techniques such as \emph{PRACTISE}~\cite{xue2015practise} could yield more accurate prediction, this example shows that very good results can be obtained with general-purpose plugins, and with little effort.

\subsection{Analysis of Job Behavior}
\label{section:casestudy2}

In the second case study, we use \daf to produce aggregated performance metrics on a per-job basis, which can be visualized to gain insight about application behavior. We combine two different plugins, showing how pipelines can be used in \daf to perform complex analyses and split computational load. The plugins discussed here represent a re-implementation of the \emph{PerSyst} framework~\cite{Guillen2014}: their purpose is to enable online visualization of job performance data for HPC users, allowing them to quickly adapt configurations and spot issues. Because of its online and user-oriented nature, this approach differs from others like the \emph{roofline model}~\cite{williams2009roofline}, which are more suited for offline analysis.

\subsubsection{Configuration} We deploy two distinct operator plugins, implementing a pipeline as described in Section~\ref{section:deployment}. The first \emph{Perfmetrics} plugin, instantiated in the Pushers, takes as input CPU and node-level data and computes as output a series of derived performance metrics, such as the \emph{cycles per instruction} (CPI), \emph{floating point operations per second} (FLOPS) or \emph{vectorization ratio}, which are useful to evaluate application performance. A second \emph{Persyst} job operator plugin is instantiated in the main Collect Agent: at each computing interval, it queries the set of running jobs on the HPC system, and for each of them it instantiates a block according to its configuration. In this case, blocks have as input one of the Perfmetrics derived metrics from all compute nodes on which the job is running. From these, the operator computes a series of job-level statistical indicators (e.g., median) as output. In the Pushers and Collect Agent, sampling and computation are performed at 1s intervals.

We executed four jobs, each on 32 \cmuc nodes and running the \emph{Kripke}, \emph{AMG}, \emph{Nekbone} and \emph{LAMMPS} proxy applications. The job runs were repeated multiple times and under different node configurations to ensure consistency. Here we focus on the CPI metric: thus, we configured the Perfmetrics plugin to have an operator with one block per CPU core, each producing as output its CPI value. Then, we use a Persyst operator, which outputs the deciles of the job-level CPI distribution at each time point, as computed by aggregating the corresponding input values for each job. Since the latter are computed per-core, each decile is aggregated from 2048 samples at a time. This allows us to gain an overall understanding of the behavior of the applications running on the HPC system, whereas the full extent of available metrics allows us to characterize their performance profile and bottlenecks.

\subsubsection{Results}

\begin{figure}[t]
\centering
\includegraphics[width=0.40\textwidth,trim={0 0 0 10}, clip=true]{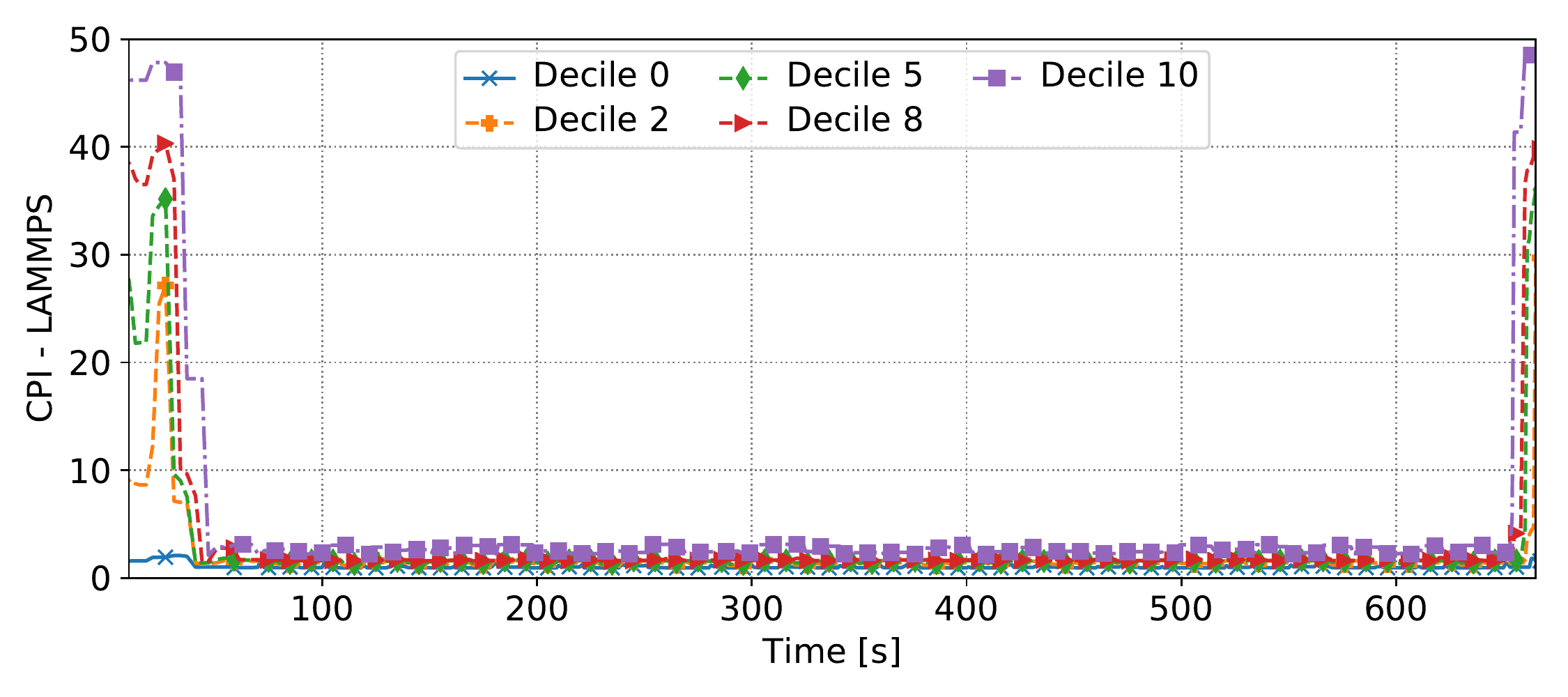} \\
\includegraphics[width=0.40\textwidth,trim={0 0 0 10}, clip=true]{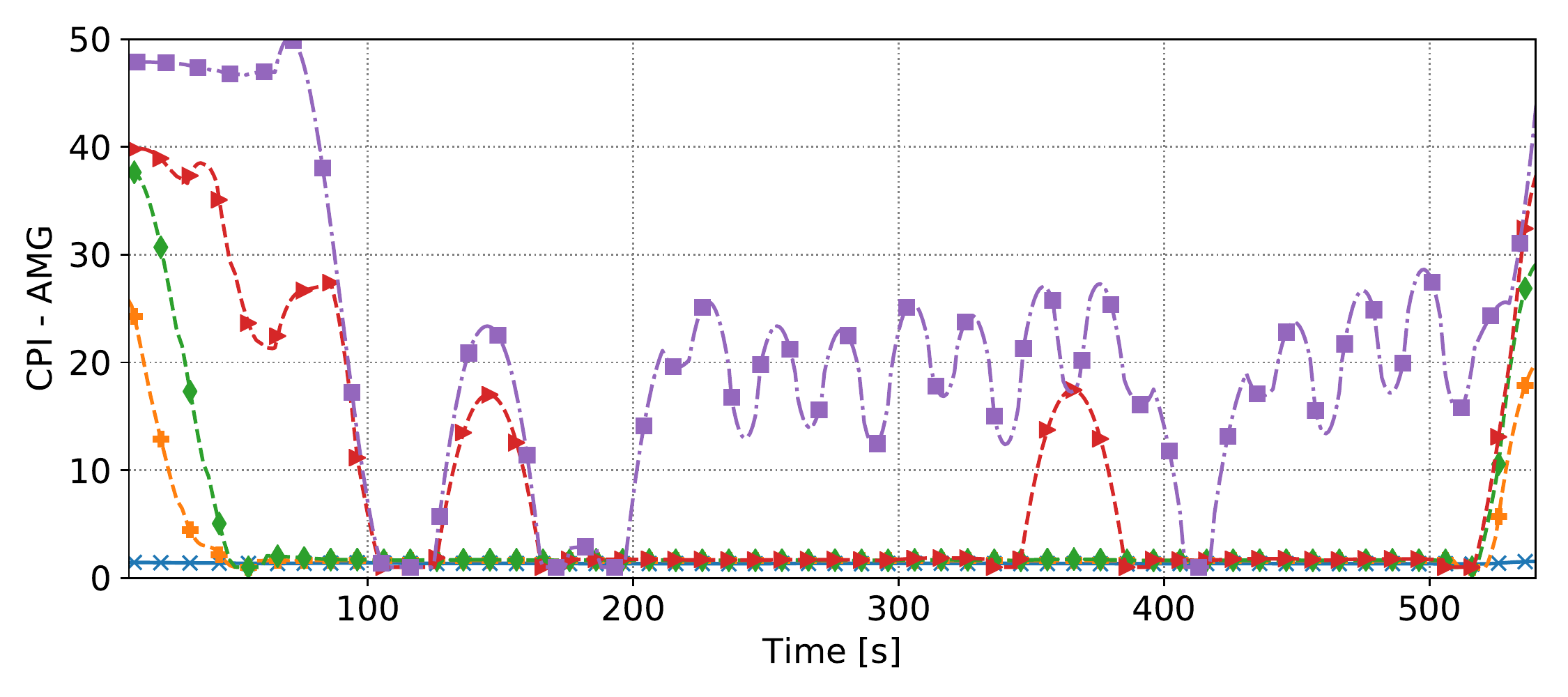} \\
\includegraphics[width=0.40\textwidth,trim={0 0 0 10}, clip=true]{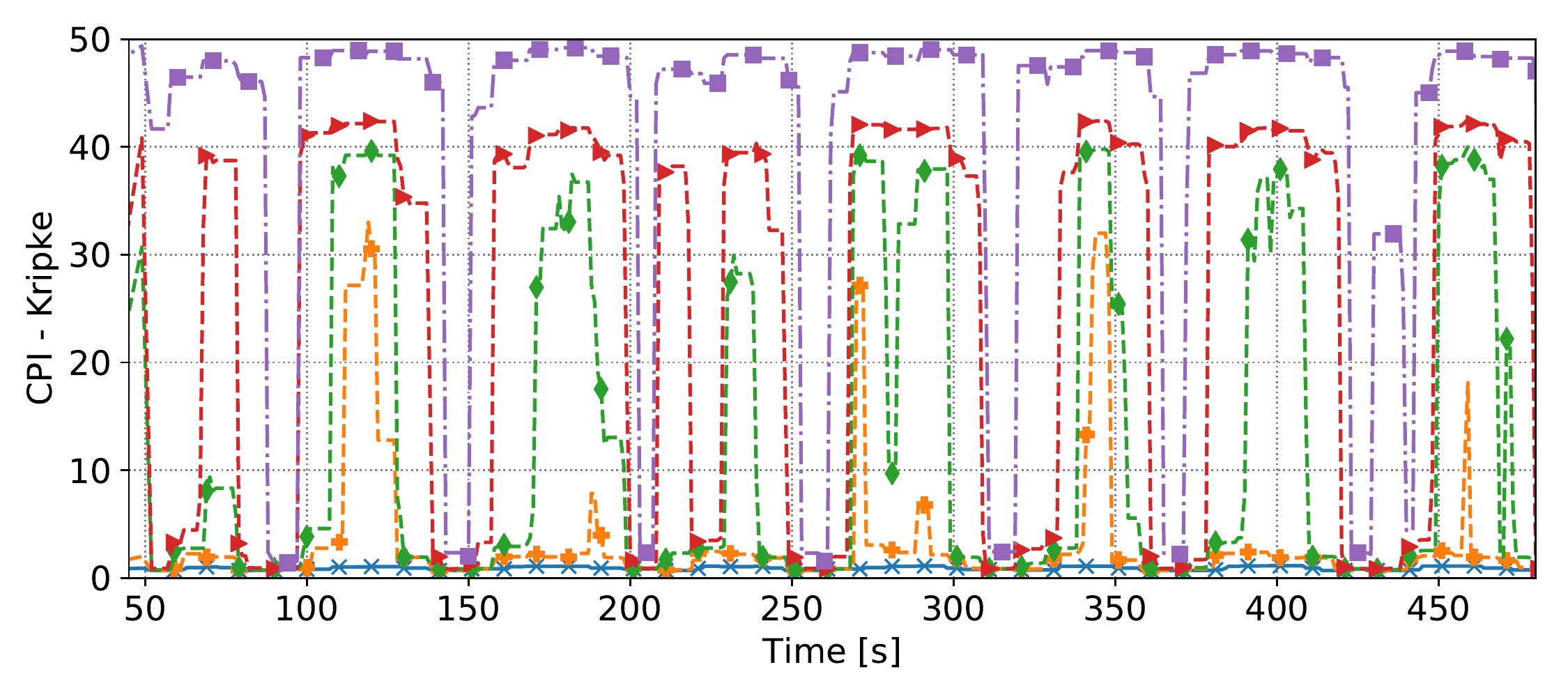} \\
\includegraphics[width=0.40\textwidth,trim={0 0 0 10}, clip=true]{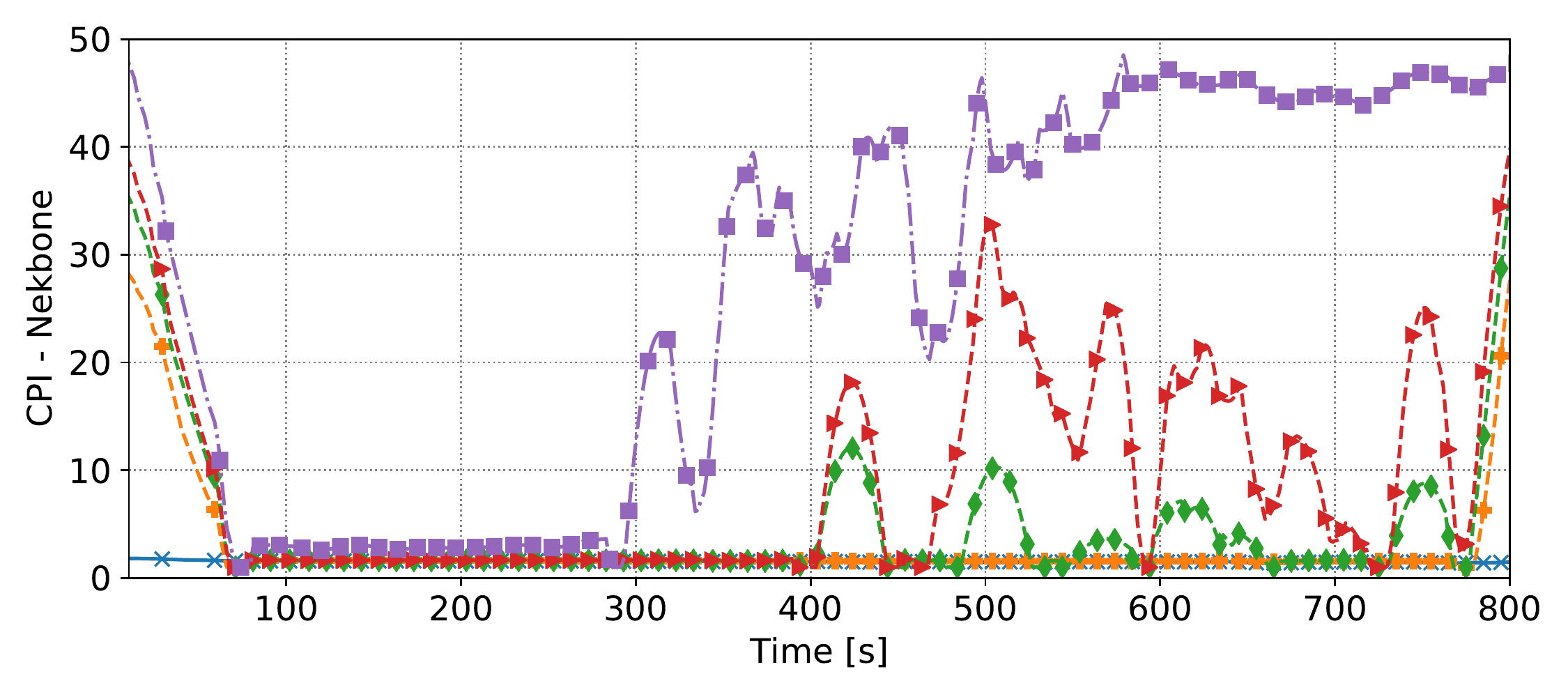}
\caption{Deciles of the aggregated per-core CPI values in function of time, for four jobs running different Coral-2 applications on \cmuc.}
\label{results:cpi}
\end{figure}

Figure~\ref{results:cpi} shows the results of our analysis: for each job, we show the time series for deciles 0, 2, 5, 8 and 10 of the aggregated per-core CPI values while running the corresponding Coral-2 codes; deciles 0, 5 and 10 correspond to the minimum, median and maximum CPI values respectively. It can be seen that the applications exhibit distinctly different behaviors depending on the underlying computational workload: LAMMPS shows low CPI values averaging at 1.6, with minimum spread in the distribution, which is due to its mostly compute-bound nature. A similar behavior can be observed with AMG, with low CPI values up to decile 5: however, deciles 8 and 10 show spikes up to CPI values of 30. As AMG is a network and memory-bound application based on fine-grained synchronization, this behavior could be caused by network latency affecting I/O, as well as load imbalance.

Kripke has a very distinctive profile: it is possible to separate each single iteration thanks to the increase and decrease in CPI values across all deciles. Similarly to AMG, Kripke is also a network and memory-bound application, and is thus characterized by relatively high CPI values. Finally, Nekbone shows the most interesting behavior: low CPI values can be observed in the first part of the application run, which is expected as Nekbone is a compute-bound application. In the second part of the run, however, the spread across deciles increases dramatically, with at least 20\% of the CPUs exhibiting higher CPI values. Our hypothesis is that, as Nekbone performs a batch of tests on increasing problem sizes, the application becomes memory-limited as soon as it grows past the 16GB-\emph{High-Bandwidth Memory} available in \cmuc nodes. This is a typical example of how visualization of performance metrics can be used to spot bottlenecks in HPC applications.

\subsection{Identification of Performance Anomalies}
\label{section:casestudy3}

For the final case study, we conduct a long-term analysis on coarse-grained monitoring data from all compute nodes in \cmuc. By applying unsupervised learning techniques, we characterize the performance of the entire HPC system and highlight variance between compute nodes, as well as identify outliers and anomalous behaviors: this can be used to automatically raise alerts to system administrators or to improve resource allocation decisions.

\subsubsection{Configuration} We use a single \emph{Clustering} plugin employing Bayesian Gaussian mixture models in the main Collect Agent. This plugin is configured to have one operator with as many blocks as compute nodes, each having as input a node's power, temperature and CPU idle time sensors, and as output a label of the cluster to which it belongs. More precisely, at every computation interval the operator computes 2-week averages for the input sensors of each block. Then, each block is treated as a data point in a three-dimensional space, and clustering is applied to them. Sampling in Pushers is performed every 10s and clustering every hour.

We adopt a Bayesian Gaussian mixture model because, unlike ordinary Gaussian mixture models, they are able to determine the optimal number of clusters from data~\cite{roberts1998bayesian}. This is useful in an online, continuous scenario, where the diverse states of an HPC system can be captured without manual tuning of the model's parameters. The number of input sensors to the clustering algorithm (and thus the number of dimensions) can be changed at will in the plugin's configuration, as well as the length of the averages' aggregation window. Since the job runtime limit is set to 2 days on \cmuc, we choose a value of 2 weeks to extract the performance profile of each node without knowledge of running jobs.

\subsubsection{Results}

\begin{figure}[t]
\centering
\includegraphics[width=0.45\textwidth,trim={50 0 0 0}, clip=true]{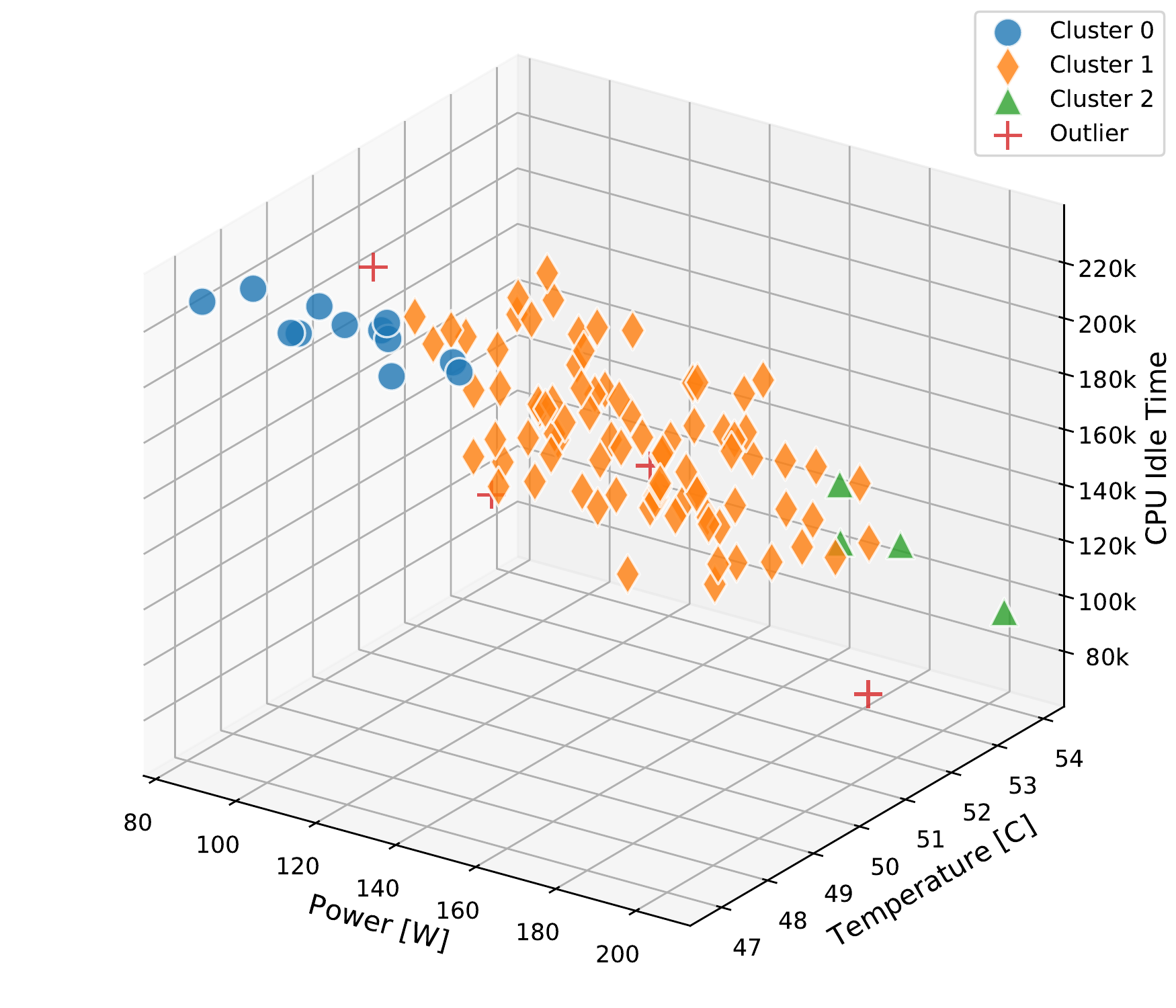}
\caption{Bayesian Gaussian mixture clustering applied to \cmuc. Each point represents a compute node in the HPC system, with the associated 2-week averages of the power, temperature and CPU idle time metrics.}
\label{results:clustering}
\end{figure}

Figure~\ref{results:clustering} shows the result of the clustering process for a single time window. The points in the scatter plot correspond to compute nodes in \cmuc, whose coordinates are their 2-week power, temperature and CPU idle time averages. First, it can be observed that the three metrics are strongly correlated, and the compute nodes describe a clear linear trend: this is expected, as a compute node will consume less power if idling, and its temperature will be lower as well. Most nodes concentrate in cluster 1 towards the center of the plot, with relatively little spread.

Despite the 2-week aggregation window we adopted, some stark differences in node behavior can still be observed: compute nodes belonging to cluster 0 have a higher CPU idle time, showing low power and temperature values accordingly. Conversely, nodes in cluster 2 were under heavier load compared to other nodes, peaking at 200W of average power consumption for a single node. While this behavior could simply be due to specific sequences of applications running on the nodes, it is more likely the symptom of a job scheduling policy that does not account for workload balance between nodes. A few points were classified as outliers when their probability was lower than a certain threshold (0.001 in our case) in the PDFs of all fitted Gaussian components, and the behavior of the corresponding nodes diverges significantly from the rest of the system. One node in particular shows a concerning trend, consuming roughly 20\% more power than other nodes with similar CPU idle time. We are currently investigating this anomaly, and plan to conduct a long-term root cause study. As shown, this type of analysis is very effective at supplying a comprehensive view of an HPC system's behavior, and can be useful to system administrators and researchers alike. Similarly, this can also be used to improve scheduling policies by considering recent node behavior.

\subsection{Performance and Scalability}
\label{subsection:performance}

\begin{figure}[t]
\centering
\captionsetup[subfigure]{}
\subfloat[Overhead in \emph{absolute} mode.]{
\includegraphics[width=0.21\textwidth,trim={0 0 100 0}, clip=true]{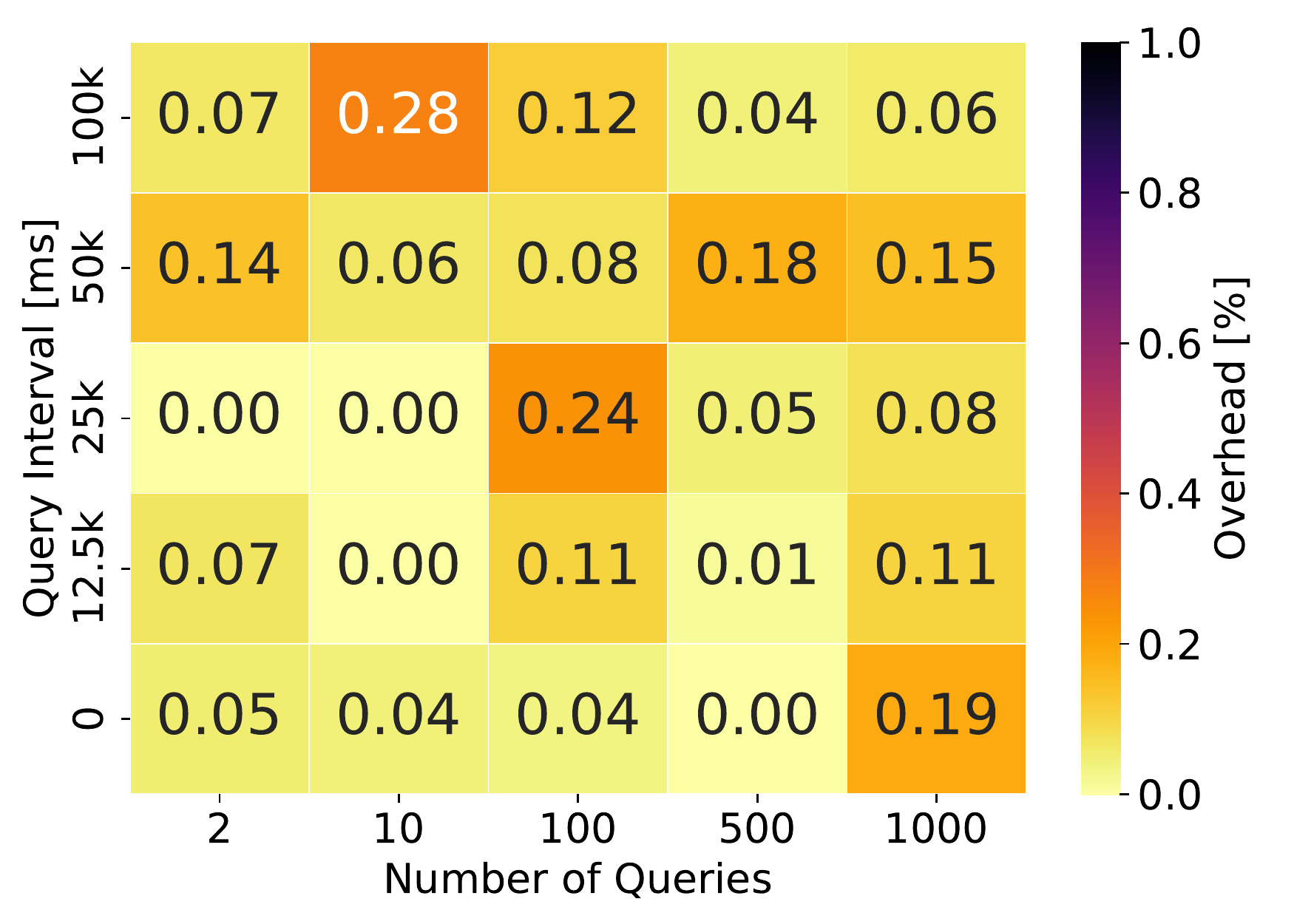}
  }
\subfloat[Overhead in \emph{relative} mode.]{
\includegraphics[width=0.25\textwidth,trim={25 0 0 0}, clip=true]{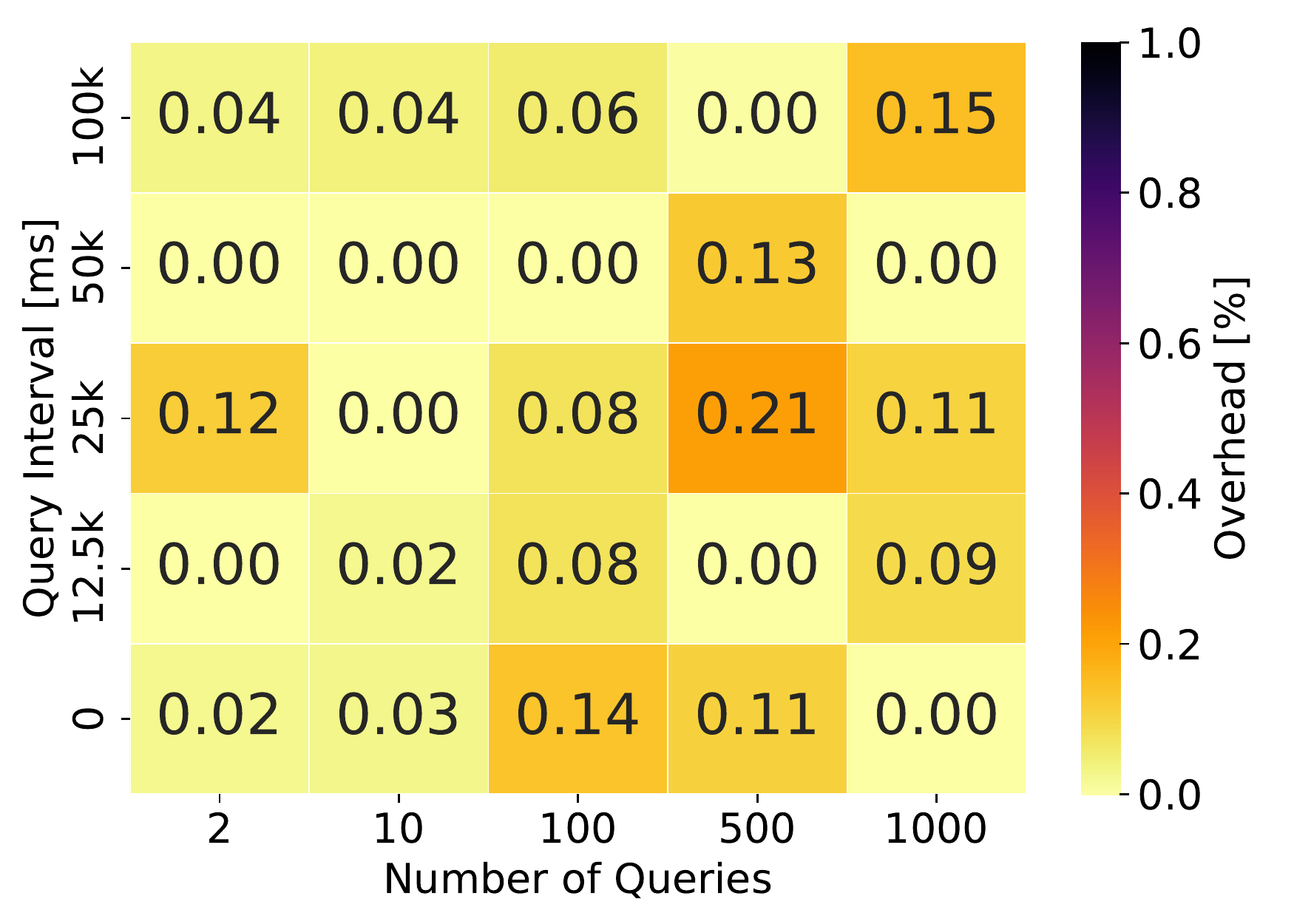}
}
\caption{Heatmaps of the query engine's median overhead at various time ranges and sensor amounts, against the HPL benchmark. A \emph{query interval} value of 0 implies that only the most recent value of each sensor is retrieved.}
\label{results:heatmaps}
\end{figure}

While the previous sections validate \daf's ability to run arbitrary ODA tasks, we now focus on the evaluation of \daf itself. The performance of \dcdb was extensively characterized in a previous work~\cite{netti2019dcdb} both on small-scale and large-scale clusters, and its overhead was found to be negligible (below 1\% for most configurations). We assume \daf to exhibit the same scaling patterns as \dcdb and hence, here we will focus on characterizing \daf's query engine component alone.

\subsubsection{Configuration}

We study the performance impact of a Pusher on the \emph{High-Performance Linpack} (HPL) benchmark~\cite{dongarra2003linpack}. In this context we use the \emph{runtime overhead}, computed as the percentage increase in execution time of HPL with a Pusher active, as opposed to running it alone. Execution times are calculated via the \emph{date} Linux command, and we instantiate a set of operators in online mode: these belong to a \emph{Querytest} plugin and simply perform a certain number of queries over the input sensors of their blocks. The input monitoring data is provided by another \emph{Tester} plugin, producing a total of 1,000 monotonic sensors with negligible overhead, so as to provide a reliable baseline. All plugins use a sampling interval of 1s and a cache size of 180s. The HPL benchmark was configured to use as many threads as physical cores on a single node, and each experiment was repeated 10 times, picking median results to ensure statistical significance and remove outliers.

\subsubsection{Performance Evaluation}

Figure~\ref{results:heatmaps} presents the results of our performance evaluation. The two heatmaps depict overhead values when varying the number of queries performed at each analysis interval, as well as the temporal range of each query, using the query engine in absolute and relative mode, respectively. Overhead is below 0.5\% in all cases, with absolute mode performing slightly worse than relative and showing higher peak overhead values: this is expected, as absolute mode employs binary search and has a higher computational complexity. Further, no clear trend can be observed when increasing the amount of queried sensor data, showing that the query engine has good scalability and minimal impact on overhead. The heatmaps are considerably noisy, likely indicating that OS noise and application variability have a larger impact on observed overhead than \daf. Average per-core CPU load of the Pusher is mostly uniform and peaks at 1.2\%. Likewise, memory usage never exceeds 25MB.

The resource footprint of \daf might be different when taking into account instantiated models and the characteristics of a production deployment. As a practical example, we measured the overhead of \daf when carrying out the case study in Section~\ref{subsection:powerusecase}, which focuses on fine-grained power consumption prediction, using the same approach discussed here. We observed that the additional overhead of performing regression on top of standard monitoring was below 0.1\% and thus negligible, showing once again the light resource footprint of \daf. Similar results were obtained for the case study in Section~\ref{section:casestudy2}, in which \daf computes derived performance metrics in-band. Here, we found the additional overhead of \daf to be always lower than 0.5\% for both the HPL benchmark and the Coral-2 applications executed on 32 nodes. In this case, the overhead increases mainly due to network interference associated with the high number of sensors, as performance metrics are computed on a per-CPU core basis. On the other hand, the case study in Section \ref{section:casestudy3} was executed out-of-band, and therefore overhead measurements are not applicable to it. We also omit quantitative comparisons with other tools: as discussed in Section~\ref{section:relatedwork}, the insular and diverse nature of existing ODA solutions renders tool comparability difficult, which is hence possible only from a qualitative standpoint.
\section{Conclusions}
\label{section:conclusions}

In this paper we have presented \daf, a framework for enabling online and holistic ODA on HPC systems, with the core objective of simplifying the instantiation of complex models for system management. Its design was conceived after an extensive literature review and requirements analysis. As a consequence, \daf is generic and can be applied to most HPC monitoring solutions: in this work, we present our implementation and integration in the \dcdb monitoring system, which is employed in our production environment. Furthermore, we adopt a novel set of logical abstractions, denoted as the block system, to partition the space of available sensors and simplify model configurations. We show that our implementation of \daf has a small resource footprint, making it suitable for applications in which latency and overhead are critical. We then present a series of case studies in the fields of runtime tuning, job analysis and performance variation: this highlights how \daf can be easily and effectively applied to many usage scenarios, on the same system, that would be otherwise difficult to implement. \daf is currently deployed to perform sensor aggregation in the CooLMUC-3 and SuperMUC-NG\footnote{\url{https://doku.lrz.de/display/PUBLIC/SuperMUC-NG}} systems at \lrz. As future work, we plan to identify additional production use cases, as well as explore solutions to simplify the management of operators and ensure high availability.

\vspace{2mm}
\textit{Acknowledgements.} This research activity has received funding from the DEEP-EST project under the EU H2020-FETHPC-01-2016 Programme grant agreement n° 754304.

%\vfill

\bibliographystyle{ACM-Reference-Format}
\bibliography{main}

\end{document}